\documentclass[final]{svjour2}
\usepackage{graphicx}
\usepackage{rotating}
\usepackage{amssymb}
\usepackage{mathptmx}
\usepackage[numbers]{natbib}
\makeatletter
\journalname{Journal of Low Temperature Physics}

\bibpunct{}{}{,}{s}{}{,}

\newcommand{\beq}{\begin{equation}}
\newcommand{\eeq}{\end{equation}} 
\newcommand{\beqa}{\begin{eqnarray}}
\newcommand{\eeqa}{\end{eqnarray}}
\newcommand{\ba}{\begin{array}}
\newcommand{\ea}{\end{array}}

\begin{document}

\newcommand{\hdblarrow}{H\makebox[0.9ex][l]{$\downdownarrows$}-}
\title{Viscosity-entropy ratio of the unitary Fermi gas 
from zero-temperature elementary excitations}
\author{Luca Salasnich and Flavio Toigo} 
\institute{Dipartimento di Fisica ``Galileo Galilei'' and CNISM, 
Universit\`a di Padova, Via Marzolo 8, 35122 Padova, Italy \\
\email{luca.salasnich@unipd.it, flavio.toigo@unipd.it}}

\date{\today}

\maketitle

\begin{abstract} 
We investigate the low-temperature behavior of the ratio between 
the shear viscosity $\eta$ and the entropy density $s$ 
in the unitary Fermi gas by using a model 
based on the zero-temperature spectra of both bosonic collective 
modes and fermonic single-particle excitations. 
Our theoretical curve of $\eta/s$ as a function of the temperature $T$ 
is in qualitative agreement with the experimental data of trapped 
ultracold $^6$Li atomic gases. We find the minimum value 
$\eta/s \simeq 0.44$ (in units of $\hbar/k_B$) 
at the temperature $T/T_F \simeq 0.27$, with $T_F$ the Fermi temperature. 
\end{abstract}

\section{Introduction}

Strongly interacting quantum many-body systems like Helium 4, 
the quark-gluon plasma and the unitary Fermi gas 
share a common feature: an extremely low viscosity 
hydrodynamics.\cite{thomas1,thomas2} These quite different 
many-body systems show a ratio of shear viscosity 
$\eta$ to entropy density $s$ which is not too far from the 
lower bound $\eta/s=\hbar /(4\pi k_B)$ predicted for a ``perfect fluid'' 
by using the anti-deSitter/conformal field theory (AdS/CFT) 
duality between certain strongly coupled 
field theories in $d=4$ space-time dimensions and weakly coupled 
string theory in $d=10$.\cite{string} 
As discussed in a recent review,\cite{thomas4} 
theoretical predictions\cite{lowt,hight} 
of the viscosity-entropy ratio for dilute and ultracold Fermi 
atoms in the unitary regime, 
where the s-wave inter-atomic scattering length $a_F$ diverges, 
are not in good agreement with the experimental data of 
the viscosity-entropy ratio measured 
in the $^6$Li atomic gas.\cite{thomas3}

In this paper we study the low-temperature behavior of $\eta/s$ by using  
a recent heuristic analysis of the shear viscosity \cite{leclair} 
and a thermodynamical model \cite{iome} of the unitary Fermi gas 
based on zero-temperature elementary excitations. 
We show that our theoretical curve for $\eta/s$ as 
a function of the temperature $T$ 
is in qualitative good agreement with the experimental data of trapped 
ultracold $^6$Li atomic gases. In particular, we find the minimum value 
$\eta/s \simeq 0.44$ (in units of $\hbar/k_B$) 
at the temperature $T/T_F \simeq 0.27$, with $T_F$ the Fermi temperature. 
Both the value and the position of this minimum are fully compatible 
with the most recent experimental determinations.\cite{thomas2,thomas3} 

In the first part of this paper we briefly review our thermodynamical 
model \cite{iome} of the unitary Fermi gas comparing it with 
experimental data \cite{jap} and Monte Carlo simulations \cite{thermo-bulgac}. 
In the second part we adopt the analysis of 
How and LeClair \cite{leclair} for the shear viscosity 
and derive from it and from our thermodynamical model \cite{iome} 
the viscosity-entropy ratio $\eta/s$. 
We then compare our curve of $\eta/s$ vs $T$ with available 
experimental data \cite{thomas3} and proposed 
theories.\cite{lowt,hight,thomas4,leclair} 

\section{Elementary excitations of the unitary Fermi gas} 

For any many-body system the weakly excited 
states, the so-called elementary excitations, 
can be treated as excitations of an ideal gas.\cite{landau,abrikosov} 
In general, these elementary excitations are the result of collective 
interactions of the particles of the system, and therefore pertain 
to the system as a whole and not to its 
separate particles.\cite{landau,abrikosov} 
For the unitary Fermi gas the mean-field extended BCS theory 
predicts the existence of fermionic 
single-particle elementary excitations characterized by 
an energy gap $\Delta$.\cite{levin}  
The inclusion of beyond-mean-field 
effects, namely quantum fluctuations of the order parameter, 
gives rise to bosonic collective excitations,\cite{levin} 
which are density waves reducing to 
the Bogoliubov-Goldstone-Anderson mode in the 
limit of small momenta.\cite{etf} 

Our effective quantum Hamiltonian \cite{iome} of the uniform unitary Fermi 
gas with two equally-populated spin components is then assumed to be: 
\beq 
{\hat H} = E_{0} + \sum_{\bf q} \epsilon_{col}(q) \ 
{\hat b}_{\bf q}^+ {\hat b}_{\bf q} + \sum_{\sigma=\uparrow,\downarrow} 
\sum_{\bf p} \epsilon_{sp}(p) \ {\hat c}_{{\bf p}\sigma}^+ 
{\hat c}_{{\bf p}\sigma} \; , 
\label{hamilt}
\eeq
where $E_{0}$ is the ground-state energy, ${\hat b}_{\bf p}^+$ and 
${\hat b}_{\bf p}$ are the bosonic 
creation and destruction operators of a collective 
excitation of linear momentum ${\bf q}$ with energy 
$\epsilon_{col}(q)$, while ${\hat c}_{{\bf p}\sigma}^+$ 
and ${\hat c}_{{\bf p}\sigma}$ are the fermionic 
creation and destruction operators of a single-particle 
excitation of linear momentum ${\bf p}$ and spin $\sigma$, with 
energy $\epsilon_{sp}(p)$.

It is now well-established \cite{levin} that the ground-state 
energy $E_0$ of the uniform unitary Fermi gas made 
of $N$ atoms in a volume $V$ is given by 
\beq 
E_0  = {3\over 5} \xi  N  \epsilon_F  \; ,   
\eeq 
with $\xi\simeq 0.4$  \cite{bertsch} and where
$\epsilon_F= \hbar^2(3\pi^2 n)^{2/3}/(2m)$ is the Fermi energy 
of a noninteracting fermi gas with density $n=N/V$. 

The exact dispersion relation of elementary 
(collective and single-particle) excitations is not fully 
known.\cite{levin} In Ref. \cite{etf} we have found 
the dispersion relation of collective elementary excitations as 
\beq
\epsilon_{col}(q) = \sqrt{c_1^2 q^2 + {\lambda\over 4m^2}q^4 } \; , 
\eeq
where 
\beq 
c_1 = \sqrt{\xi\over 3}\ v_F  \; , 
\label{disp}
\eeq
is the zero-temperature first sound velocity, 
with $v_F=(\hbar/m)(3\pi^2 n)^{1/3}$ the 
Fermi velocity of a noninteracting Fermi gas. Notice that 
the term with $\lambda$ takes into account the increase 
of kinetic energy due to spatial variations 
of the density\cite{etf,kim,manini05,sala-josephson,sala-new,recent,cristina}. 
For the purposes of the present paper, by fixing $\xi=0.42$, 
i.e. the value given by the Monte Carlo prediction 
for a uniform gas of Astrakharchik {\it et al.},\cite{astra} 
we find that the best agreement with Monte Carlo 
data is obtained with $\lambda =0.25$. 
 
The collective modes  describe correctly only 
the low-energy density oscillations of the system while
at higher energies one expects the appearence of 
fermionic single-particle excitations starting 
from the threshold above which there is the breaking 
of Cooper pairs.\cite{levin,thermo-bulgac,magierski1} 
At zero temperature these single-particle elementary excitations 
can be written as 
\beq 
\epsilon_{sp}(p) = 
\sqrt{\big({p^2\over 2m} - \zeta \epsilon_F \big)^2 + \Delta_0^2}  
\eeq
where $\zeta$ is a parameter which takes into account the interaction 
between fermions ($\zeta \simeq 0.9$ according to recent Monte Carlo results 
\cite{magierski1}) with $\epsilon_F$ the Fermi energy of the ideal Fermi gas.  
$\Delta_0$ is the zero-temperature gap parameter 
with $2\Delta_0$ the minimal energy to break a Cooper 
pair.\cite{levin} 
Notice that the gap energy $\Delta_0$ of the unitary Fermi gas 
at zero-temperature has been calculated with Monte Carlo 
simulations \cite{magierski1,carlson} and found to be
$\Delta_0=\gamma \epsilon_F $, with $\gamma \simeq 0.45$.  

\section{Thermodynamics of the unitary Fermi gas} 
  
At very low temperature 
the thermodynamic properties of the superfluid unitary Fermi gas can 
be obtained from the collective spectrum 
and considering it as an ideal Bose gas of  
elementary excitations \cite{landau} with the bosonic 
distribution 
\beq 
f_B(q) = 
\langle {\hat b}_{\bf q}^+{\hat b}_{\bf q} \rangle  
= {1\over e^{\epsilon_{col}(q)/k_BT} - 1} \; , 
\eeq
where $\langle {\hat A} \rangle = 
Tr[ {\hat A} e^{-{\hat H}/k_BT}]/Tr[e^{-{\hat H}/k_BT}]$ 
is the thermal average of the operator ${\hat A}$ 
with $T$ the absolute temperature and $k_B$ is the Boltzmann constant.
\cite{kerson} As $T$ increases also the fermionic single-particle 
excitations become important. Thus there is also the effect of 
an ideal Fermi gas of single-particle excitations  
with the fermionic distribution 
\beq 
f_F(p) = 
\langle {\hat c}_{{\bf p}\sigma}^+{\hat c}_{{\bf p}\sigma} \rangle  
= {1\over e^{\epsilon_{sp}(p)/k_BT} + 1} 
\label{fermi-dirac}
\; ,   
\eeq 
which is spin independent. 

The Helmholtz free energy $F$ of any thermodynamic system 
is given by 
\beq 
F = -k_B T \ln{\cal Z} \; ,  
\eeq
where 
\beq 
{\cal Z} = Tr[e^{-{\hat H}/k_BT}] \; , 
\eeq
is the partition function of the system.\cite{kerson} 
Using Eq. (\ref{hamilt}) the free energy of our unitary Fermi gas  
can be written as $F = F_0+F_{col}+F_{sp}$, 
where $F_0$ is the free energy of the ground-state, $F_{col}$ 
is the free energy of the bosonic collective excitations and 
$F_{sp}$ the free energy of fermionic single-particle excitations. 
The Helmholtz free energy $F_0$ of the uniform ground state coincides with 
the zero-temperature internal energy $E_0$ and is given by 
\beq 
F_0 = {3\over 5} \xi  N  \epsilon_F  \; ,   
\eeq 
where $N$ is the number of atoms of the uniform system in a volume $V$. 
The free energy $F_{col}$ of the collective excitations   
is instead given by \cite{landau} 
\beq 
F_{col} = k_BT \sum_{\bf q} 
\ln{\left[ 1 - e^{-\epsilon_{col}(q)/(k_BT)} 
\right]} \; , 
\eeq
while the free energy $F_{sp}$ due to the single-particle 
excitations is 
\beq 
F_{sp} = - 2 \ k_B T \sum_{\bf p} 
\ln{\left[ 1 + e^{-\epsilon_{sp}(p)/(k_BT)} 
\right]} \; , 
\eeq
where the factor $2$ is due to the two spin components. 
As previously discussed, the total Helmholtz free energy 
$F$ of the low-temperature 
unitary Fermi gas can be then written as $F_0+F_{col}+F_{sp}$, namely 
\beq 
F = N \epsilon_F \Phi\left({T\over T_F}\right) \; , 
\label{free} 
\eeq 
where $\Phi(x)$ is a function of the scaled temperature $x=T/T_F$, 
with $T_F=\epsilon_F/k_B$, given by 
\beqa 
\Phi(x) &=& {3\over 5}\xi + {3\over 2} x \int_0^{+\infty} 
\ln{\left[ 1 - e^{-{\tilde \epsilon}_{col}(u)/x} \right]}u^2 du 
\nonumber
\\
&-& 3 x \int_0^{+\infty} 
\ln{\left[ 1 + e^{-{\tilde \epsilon}_{sp}(u)/x}\right]} u^2 du 
\; . 
\label{free-scaled} 
\eeqa
Here the discrete summations have been replaced by integrals, 
moreover we set ${\tilde \epsilon}_{col}(u)=\sqrt{u^2(\lambda u^2+4\xi/3)}$  
and ${\tilde \epsilon}_{sp}(u)=\sqrt{(u^2-\zeta)^2+\gamma^2}$. 

From the Helmholtz free energy $F$ we can immediately calculate the 
chemical potential $\mu$, through 
\beq 
\mu = \left({\partial F \over \partial N}\right)_{T,V} \; . 
\eeq
obtaining
\beq
\mu = \epsilon_F \Big[ {5\over 3} \Phi \left({T\over T_F}\right) 
-{2\over 3} {T\over T_F} \Phi'\left({T\over T_F}\right) \Big] 
\; , 
\label{chemical}
\eeq
where $\Phi'(x)={d\Phi(x)\over dx}$ and one recovers 
$\mu_0=\xi \epsilon_F$ in the limit of zero-temperature. 

The entropy $S$ is related to the free energy $F$ by the formula
\beq 
S = - \left({\partial F\over \partial T}\right)_{N,V} \; , 
\eeq
from which we get 
\beq 
S = - N k_B \Phi'\left({T\over T_F}\right) \; .    
\label{entropy}
\eeq 
In addition, the internal energy $E$, given by 
\beq 
E = F + T S \; , 
\eeq 
can be written explicitly as 
\beq 
E = N \epsilon_F 
\left[ \Phi\left({T\over T_F}\right) - {T\over T_F} 
\Phi'\left({T\over T_F}\right) \right] \; .   
\label{internal}
\eeq

\begin{figure}
\begin{center}
\includegraphics[width=0.8\linewidth,clip]{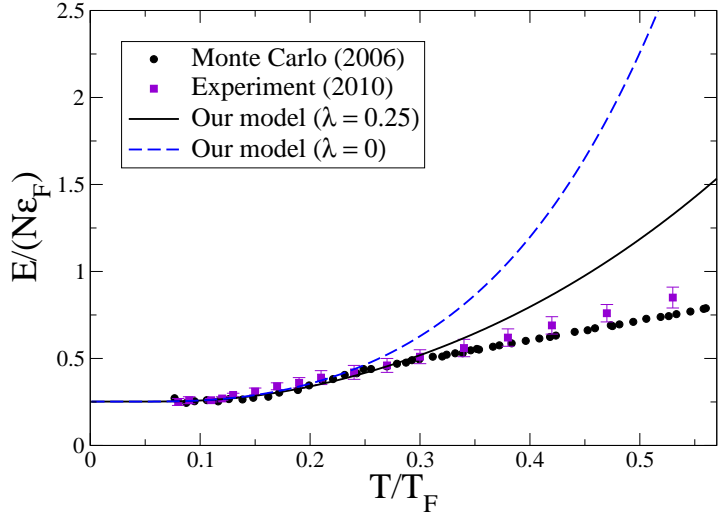}
\end{center}
\caption{(Color online). Scaled internal energy 
$E/(N\epsilon_F)$ as a function 
of the scaled temperature $T/T_F$. 
Filled circles: Monte Carlo simulations \cite{thermo-bulgac}. 
Open squares with error bars: experimental data of 
Horikoshi {\it et al.} \cite{jap}. 
Solid line: our model, i.e. Eq. (\ref{internal}) 
and Eq. (\ref{free-scaled}), with $\lambda=0.25$. 
Solid line: our model, i.e. Eq. (\ref{internal}) 
and Eq. (\ref{free-scaled}), with $\lambda=0$. 
Other zero-temperature parameters of elementary 
excitations: $\xi=0.42$, $\zeta=0.9$, and $\gamma=0.45$.} 
\label{f1}
\end{figure}

It is interesting to compare our model with other theoretical 
approaches and also with the available experimental data. 
In Fig. \ref{f1} we report the data of internal energy $E$ 
obtained by Bulgac, Drut and Magierski \cite{thermo-bulgac} with 
their Monte Carlo simulations (filled circles) of the atomic unitary 
gas. We insert also the very recent experimental data of 
Horikoshi {\it et al.} \cite{jap} for the unitary Fermi gas of $^6$Li atoms 
but extracted from the gas under harmonic confinement (filled squares 
with error bars). In the figure we include the results of 
our model, that is given by 
Eqs. (\ref{internal}) and (\ref{free-scaled}) 
with both $\lambda=0.25$ (solid line) and $\lambda=0$ (dashed line). 
The figure shows that in our model the gradient 
term, proportional to $\lambda$, plays a marginal role up 
to $T/T_F\simeq 0.25$. 
Above $T/T_F\simeq 0.25$, however, our results with $\lambda=0.25$ 
shows a better agreement with both Monte Carlo and experimental 
data than those with $\lambda=0$. We stress that the gradient term  
is essential to describe accurately the zero-temperature 
surface effects of a trapped system, in particular 
with a small number of atoms, 
where the Thomas-Fermi (i.e. $\lambda=0$) 
approximation fails \cite{etf}. The value $\lambda=0.25$ gives the best fit 
of the Monte Carlo energy as a function of the particle number 
for $\xi=0.42$ (see Ref.\cite{etf,iome} for details). 

Our model does not show a phase transition. Nevertheless, 
the results of Fig. \ref{f1} strongly shows that our model works 
quite well not only in the superfluid regime, but also slightly above the 
critical temperature ($T_c/T_F\simeq 0.15$) suggested by 
two theoretical groups.\cite{magierski1,troyer} 
This finding is not fully surprising. 
In presence of a pseudo-gap region, the temperature-dependent 
gap $\Delta(T)$ of single-particle elementary excitations can be written as 
$\Delta(T) = \Delta_{sc}(T)+\Delta_{pg}(T)$, where $\Delta_{Sc}(T)$ is the 
superconducting gap and $\Delta_{pg}(T)$ is the pseudogap.\cite{levin} 
At $T_c$ the superconducting gap $\Delta_{sc}(T)$ 
goes to zero, i.e. $\Delta_{sc}(T_c)=0$, but the pseudo-gap 
$\Delta_{pg}(T)$ remains finite and it becomes zero only at the 
higher temperature $T^*$.\cite{levin} For further details 
on the comparision between our model and other theories 
see Ref. \cite{iome}. 

\section{Shear viscosity from thermodynamics} 

A first principle calculation of the shear viscosity is beyond the 
scope of the present work and we adopt the heuristic 
analysis of How and LeClair\cite{leclair} 
to write it in terms of the scaled free energy $\Phi(x)$ 
and its first derivative $\Phi'(x)$. 
The shear viscosity $\eta$ can be estimated by using 
the formula\cite{lebellac} 
\beq 
\eta = {1\over 3}\ n \ m \ \bar{v} \ l_{m} \; , 
\eeq
where $n$ is the total number density of the fluid, $m$ is the mass 
of each particle in the fluid, $\bar{v}$ is the average velocity 
of particles, and $l_{m}$ is the length of the mean free path. 
The mean free path is written as 
\beq 
l_{m} = {1\over n \sigma} \; , 
\eeq
where $\sigma$ is a suitable transport cross-section.\cite{lebellac} 

\begin{figure}
\begin{center}
\includegraphics[width=0.8\linewidth,clip]{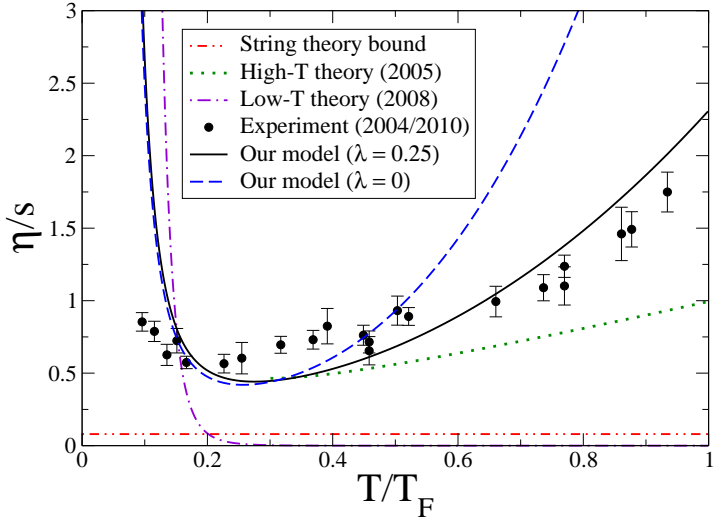}
\end{center}
\caption{(Color online). Viscosity-entropy ratio 
$\eta/s$ (in units of ${\hbar}$/${k_B}$) 
as a function of the scaled temperature $T/T_F$. 
Dot-dot-dashed line: the bound $\eta/s=1/(4\pi)$ predicted by field theory 
models inspired by string theory.\cite{string} 
Filled circles with error bars: experimental data.\cite{thomas3,thomas4} 
Dot-dashed line: low-temperature prediction of Ref. \cite{lowt}. 
Dotted line: high-temperatue prediction of Ref. \cite{hight}. 
Solid line: our model with $\lambda=0.25$. 
Dashed line: our model with $\lambda=0$. 
Other zero-temperature parameters of elementary 
excitations: $\xi=0.42$, $\zeta=0.9$, and $\gamma=0.45$.} 
\label{f2}
\end{figure}

For the unitary gas with two-spin-component fermions, the cross-section 
is given by
\beq 
\sigma = {4\pi\over |{\bf k}_1-{\bf k}_2|^2} \; , 
\label{bunga}
\eeq
where ${\bf k}_1-{\bf k}_2$ is the relative wave number of two 
colliding fermions with opposite spin.\cite{landau} 
The average velocity $\bar{v}$ of fermions 
can be related to the relative wave number $|{\bf k}_1-{\bf k}_2|$ 
by the formula\cite{leclair} 
\beq 
\bar{v} = \sqrt{2} {\hbar\over m} 
\langle |{\bf k}_1-{\bf k}_2|^2\rangle^{1/2} \; . 
\eeq
In fact, $\langle |{\bf k}_1 - {\bf k}_2|^2 \rangle = 
\langle k_1^2 + k_2^2 - 2 {\bf k}_1 \cdot {\bf k}_2 \rangle = 
2 \bar{k}^2$, because $\langle {\bf k}_1 \cdot  {\bf k}_2 \rangle 
=0$ and $\bar{k}^2 = \langle k_1^2 \rangle = \langle k_1^2 \rangle$, 
and finally $\bar{v}=(\hbar/m) \bar{k}$.\cite{nota-leclair} 
In this way the shear viscosity becomes 
\beq 
\eta = {m^3 \bar{v}^3 \over 6\pi \hbar^2} \; . 
\label{eta}
\eeq
The average velocity $\bar{v}$ can be estimated by 
imposing that the average kinetic energy is 
equal to the internal energy per particle,\cite{leclair} namely 
\beq 
{1\over 2} m \bar{v}^2 = {E\over N} \; .   
\label{kinetic}
\eeq 
By using Eq. (\ref{eta}) with $\bar{v}$ given by Eq. (\ref{kinetic}) 
and $E$ given by Eq. (\ref{internal}), the shear viscosity reads 
\beq 
\eta = n \hbar {\pi\over 2} \left( \Phi({T\over T_F}) - {T\over T_F} 
\Phi'({T\over T_F}) \right)^{3/2} \ . 
\eeq
Notice that for $T\to 0$, the viscosity $\eta$ goes to a constant 
value because also $\sigma$ goes to a constant $\simeq k_F^{-2}$. 
This is in agreement with recent experimental results on the universal 
spin diffusion in a strongly interacting Fermi gas.\cite{roati} 
Finally, by considering Eq. (\ref{entropy}) for the entropy of the 
unitary Fermi gas, we find that the viscosity-entropy ratio is 
given by 
\beq 
{\eta\over s} = - {\hbar\over k_B} {\pi\over 2} 
{\left( \Phi({T\over T_F}) - {T\over T_F} 
\Phi'({T\over T_F}) \right)^{3/2} \over \Phi'({T\over T_F}) } \; ,  
\label{ratio}
\eeq
where $s=S/V$ is the entropy density, 
i.e. the entropy per unit of volume. This formula gives the viscosity-entropy 
ratio in terms of the scaled free energy $\Phi(x)$ and its first derivative 
$\Phi'(x)$. For $T\to 0$ 
Eq. (\ref{ratio}) gives $\eta/s=+\infty$. This divergence of 
$\eta/s$ is a consequence of Eqs. (\ref{eta}) and (\ref{kinetic}) 
which impose, as previously stressed, a small 
but finite viscosity $\eta$ while the entropy density $s$ goes to zero. 

In Fig. \ref{f2} we plot experimental data of the ratio 
$\eta/s$ (filled circles with error bars). These data have been 
obtained by the group of Thomas\cite{thomas3} from the damping 
of radial breathing mode of the atomic cloud, and then elaborated 
by Sch\"afer and Chafin\cite{thomas4} with an 
energy-to-temperature calibration and averaging 
the local ratio $\eta/s$ over the trap size. 
In the figure we insert also the bound 
from string theory (dot-dot-dashed line), 
the low-temperature prediction of Rupak and Sch\"afer \cite{lowt} 
(dot-dashed line), and the high-temperature prediction 
of Bruun {\it et al.}\cite{hight} (dotted line). 
We plot also the results obtained with our model, Eq. (\ref{ratio}) 
with Eq. (\ref{free-scaled}), for $\lambda=0.25$ (solid line) 
and $\lambda=0$ (dashed line). The figure shows that 
our model is in good qualitative 
agreement with the experimental data up to $T/T_F\simeq 0.4$. 
Both with $\lambda=0.25$ (solid line) and $\lambda=0$ (dashed line) 
our model shows a minimum for $\eta/s\simeq 0.44$ at $T/T_F\simeq 0.27$. 
Notice that the solid curve ($\lambda =0.25$) gives a reasonable 
agreement up to $T/T_F\simeq 0.9$. 

We observe that the curve of $\eta/s$ vs $T/T_F$ obtained by How 
and LeClair, \cite{leclair} on the the basis 
of their version of Eq. (\ref{ratio}) but with 
a very different procedure to calculate the scaled free energy 
$\Phi(x)$, does not seem to increase 
as $T/T_F$ goes to zero. Actually, a very recent calculation\cite{levin2} 
of the shear viscosity from current-current correlation functions 
suggests that $\eta/s$ at low $T$ becomes small rather than exibiting 
the upturn. Nevertheless, the obtained theoretical values\cite{levin2} 
of $\eta/s$ appear quite large with respect to the experimental ones. 

\section{Conclusions} 

We have described the elementary excitations of the 
unitary Fermi gas as made of collective bosonic excitations  
and fermionic single-particle ones. 
We have obtained an analytical expression for 
the Helmholtz free energy, showing that it is reliable to 
study the low-temperature thermodynamics of 
the unitary Fermi system up the critical temperature 
of the superfluid phase transition. 
By using this free energy and simple scaling arguments 
we have derived the viscosity-entropy ratio $\eta/s$ as a function 
of the scaled temperature $T/T_F$. Contrary to other predictions, 
our curve of $\eta/s$ vs $T/T_F$ 
is in reasonable agreement with the available experimental data. 

\begin{acknowledgements}
The authors acknowledge Aurel Bulgac, Joaquin Drut, Piotr Magierski, 
Munekazu Horikoshi and Thomas Sch\"afer 
for making available their data. LS thanks Aurel Bulgac, 
Pietr Magierski, Thomas Sch\"afer and Andr\'e LeClair 
for useful suggestions. 

\end{acknowledgements}

\pagebreak

\end{document}